\input phyzzx
\input epsf
\def\dL{{\dot L}}
\def\dR{{\dot R}}
\def\dM{{\dot M}}
\def\bP{{\overline P}}
\def\bR{{\overline R}}
\def\dtau{{\dot \tau}}

\vskip 0.25in
\centerline{\seventeenrm Canonical Quantization and the Statistical Entropy } 
\centerline{\seventeenrm of the Schwarzschild Black Hole}
\vskip 1.0in
\centerline{{\caps Cenalo Vaz}\footnote{\dagger}{Email: cvaz@haar.pha.jhu.edu.}
\footnote{\ddagger}{On leave of absence from the Universidade do Algarve, Faro, 
Portugal.
Email: cvaz@ualg.pt}}
\centerline{\it Department of Physics and Astronomy}
\centerline{\it The Johns Hopkins University}
\centerline{\it Baltimore, MD 21218.}
\vskip 1.0in

\centerline{\caps Abstract}
\vskip 0.5in

\noindent The canonical quantization of a Schwarzschild black hole yields a picture of 
the black hole that is shown to be equivalent to a collection of oscillators whose 
density of levels  is commensurate with that of the statistical bootstrap model. Energy 
eigenstates of definite parity exhibit the Bekenstein mass spectrum, $M \sim 
\sqrt{\cal N} M_p$, where ${\cal N} \in {\bf N}$.  From the microcanonical ensemble, we 
derive the statistical entropy of the black hole by explicitly counting the microstates 
corresponding to a macrostate of fixed total energy.
\vskip 1in
\noindent PACS: 04.60.Ds, 04.70.Dy
\vfill\eject

\noindent{\bf 1. Introduction}

It has been recognized for some time that black holes behave as thermodynamic objects with 
a characteristic temperature and entropy and that these quantities are inherently quantum 
mechanical in nature.${}^{1,2}$ This makes a clear understanding of the origins of black 
hole thermodynamics in terms of  statistical principles one of the more interesting open 
problems, because a microscopic description of the black hole entropy requires 
a quantum theory of gravity so that detailed first-principle investigations concerning 
the black hole entropy should contribute toward a better understanding of how such a 
theory may be constructed and interpreted. The earliest attempt at a microscopic theory 
of black holes was due to  Bekenstein${}^{3}$, who concluded that the horizon area is 
the analog of an adiabatic invariant in mechanics. He then invoked the 
Christodoulou-Ruffini process${}^{4}$ and the Bohr-Sommerfeld quantization rules to 
argue that the eigenvalues of the horizon area operator, ${\hat{\cal A}}$, of the black 
hole must be equally spaced, ${\cal A}_n  \sim n l_p^2$. Dividing  the horizon in cells 
of  Planck area which get added one at a time, and assuming that each cell has the same 
(small) number of  states, say $k$,  he was able to derive the area law of black hole 
thermodynamics by estimating the number of microscopic states to be $\Omega \approx k^n$. 

Bekenstein's hypothesis has led to many interesting attempts to derive 
the area quantization law and the black hole entropy from first principles. These attempts 
have mostly taken either the loop (or canonical) quantum gravity${}^{5}$ approach or 
the string theory${}^{6}$ approach.  The ability to reproduce the Bekenstein-Hawking 
entropy from a microscopic counting of states must be considered a measure of the 
success of a candidate quantum theory of gravity. Thus, the successful derivation 
of the Bekenstein-Hawking entropy, for example, from a microcanonical ensemble 
of  $D-$brane states${}^{7}$  is generally considered a 
triumph for string  theory and 
a necessary condition for string theory to be a convincing 
candidate for a theory of  
quantum gravity.

An approach to the entropy problem that makes explicit reference neither to string theory 
nor to canonical quantum gravity is to count the number of  states for a conformal field 
theory corresponding to the asymptotic symmetries of the black hole or to the symmetries of 
its horizon.  This mechanism for calculating the number of microstates was recently proposed 
by Strominger${}^{8}$ for the BTZ black hole.${}^{9}$ Strominger used a result of 
Brown and Henneaux${}^{10}$ which says that the asymptotic symmetry group of 
AdS${}_3$ is generated by two copies of  the Virasoro algebra with central charge $3l/2G$, 
and combined this result with Cardy's formula${}^{11}$ to compute the asymptotic growth 
of states for this conformal field theory. A semi-classical analysis of a hot black hole 
suggests that the horizon is thermally oscillating while maintaining a fixed area. 
Recently, Carlip${}^{12}$ has argued that Strominger's result is more 
generic than was originally believed, because the algebra of surface deformations of any 
black hole in any dimension contains a Virasoro algebra consisting of deformations that 
leave the horizon fixed. Cardy's formula once again yields the correct Bekenstein-Hawking 
entropy. In this approach, the horizon is treated as a boundary and all of the relevant 
degrees of freedom of the black hole are assumed to lie on it. The states themselves are 
not explicitly displayed.

The results from loop quantum gravity have been no  less dramatic. Here spin network 
states are used as a complete, orthonormal basis in the Hilbert space and loop microstates 
are explicitly counted using techniques developed in ref.[13]. In one approach${}^{14}$,  
loop quantum gravity is used to compute the microstates of the horizon, holding
the area fixed. In another approach, one analyzes the classical theory outside, treating the 
horizon as a boundary. Quantization yields surface states which are counted by an effective 
Chern-Simons theory on the boundary.${}^{15}$  Both approaches yield an entropy proportional 
to the area of the horizon, but the proportionality constant is a free, finite and 
dimensionless parameter, not determined by the theory and the horizon area eigenvalues are 
not equally spaced. 

In this article we take a midi-superspace approach to the canonical quantization 
of the Schwarzschild black hole and show that it leads naturally to Bekenstein's 
mass quantization law (equally spaced area eigenvalues) as well as 
to the area law of black hole entropy as computed from a genuine 
microcanonical ensemble. The quantization leads to an amusing picture of a black 
hole which is akin to statistical bootstrap models for hadrons, whose statistical 
mechanics was studied many years ago by Frautschi${}^{16}$ and Carlitz${}^{17}$ . 
We will use similar techniques to study the statistical 
properties of the Schwarzschild black hole, but our starting point will be a recent 
solution${}^{18}$ of the Wheeler-DeWitt equation in terms of variables first 
introduced by Kucha\v r and Brown.${}^{19,20}$
\vskip 0.25in

\noindent{\bf 2. Quantization}

As we have previously shown${}^{18}$, combining the Hamiltonian reduction of 
spherical geometries due to Kucha\v r${}^{19}$ and the coupling to dust as proposed 
by Brown and Kucha\v r${}^{20}$ allows for the derivation of  a simple, decoupled  
(Wheeler-DeWitt) equation describing the Schwarzschild black hole. The non-rotating 
dust is introduced in such a way that its role is only  as a time keeper.  Consider 
the gravity-dust system,
$$S~~ =~~ -~ {1 \over {16\pi}} \int d^4x \sqrt{-g} {\cal R}~  -~ {1 \over 2} \int 
d^4 x \sqrt{-g} \epsilon(x) \left[ g_{\alpha \beta} U^\alpha U^{\beta} + 1 \right], 
\eqno(2.1)$$
in the general spherically symmetric spacetime,
$$ds^2~~ =~~ N^2 dt^2~ -~ L^2 (dr - N^r dt)^2~ -~ R^2 d\Omega^2, \eqno(2.2)$$
where $N(t,r)$ and $N^r(t,r)$ are respectively the lapse and shift functions, 
$R(t,r)$ is the physical radius or curvature coordinate, $\epsilon(t,r)$ is 
the density of the collapsing dust in its proper frame, ${\cal R}$ is the scalar 
curvature and $U^\alpha$ are the components of the dust velocity. The action, 
$S$, in (2.1), may be be recast into the form 
$$\eqalign{S~~ =~~ \int dt dr &\left[ P_L \dL~  +~  P_R \dR~  +~ P_\tau \dtau~  
-~  N H~  -~  N^r H_r \right]\cr &+~~ surface~ terms,\cr} \eqno(2.3)$$
where we have introduced, following Brown and Kucha\v r,${}^{20}$ the dust 
proper time variable, $\tau$, which in general will serve as extrinsic time. 
$P_L$ and $P_R$ are the momenta conjugate to $L$ and $R$ respectively, and 
the super Hamiltonian, $H$, and super momentum, $H_r$, are respectively given 
by
$$\eqalign{H~~ =~~ -~ \left[{{P_L P_R} \over R}~  -~  {{L P_L^2}
\over {2R^2}}\right]~  &+~ \left[ -~ {L \over 2}~  -~  {{R'}^2 \over {2L}}~  +~  
\left({{RR'} \over L}\right)'\right]\cr  &+~ P_\tau\sqrt{1~ +   {\tau'}^2/L^2}
\cr}\eqno(2.4)$$
and 
$$H_r~~ =~~ R' P_R~  -~  L P_L'~  +~  \tau' P_\tau,\eqno(2.5)$$
where the prime denotes a derivative with respect to the ADM label coordinate $r$. 
The constraints in the above form do not ``decouple'' and are very difficult 
to resolve as they stand. However, from the general system in (2.2), Kucha\v r${}^{19}$ 
showed how one can pass by a canonical transformation to a new canonical chart with 
coordinates $M$ and $R$ together with their conjugate momenta, $P_M$ and $\bP_R$, 
where $M$ is the Schwarzschild ``mass'' and $R$ is the curvature coordinate. 
In this system the constraints are greatly simplified and the phase space 
variables have immediate physical significance. The canonical transformation is 
well-defined as long as the metric obeys standard fall-off conditions${}^{19}$ and, 
as long as these fall-off conditions are obeyed, the surface action can be 
recast in the form
$$surface~ terms~~ =~~ \int dt \left[ \pi_+ \dtau_+~  +~  \pi_- \dtau_-~  -~  
N_+ C_+~  -~  N_- C_- \right],\eqno(2.6)$$
where $\tau_\pm$ are the proper times measured on the parametrization clocks 
at right (left) infinity. The constraints $C_\pm = \pm \pi_\pm + M_\pm$ 
identify their conjugate momenta as the mass at right (left) infinity. 
In terms of the new variables the entire action, along with the surface term is
$$S~~ =~~ \int dt \int dr \left[\bP_M \dM~  +~  \bP_R \dR~  +~  \bP_\tau \dtau~  
-~  NH~  -~  N^r H_r\right],\eqno(2.7)$$
where $\bP_M = P_M - \tau'$, $\bP_\tau = P_\tau + M'$ and $P_M$ and $P_\tau$ are
the original Brown-Kucha\v r variables. The transformations leading up to these 
variables may be found in ref.[19]. In passing to the transformed momenta, 
$\bP_\tau$ and $\bP_M$, we have made a canonical transformation${}^{18}$ 
generated by $M\tau'$, which effectively absorbs the surface terms. (Thus, we have 
implicitly fixed the dust proper time to coincide at infinity with the 
parametrization clocks.) 

The super Hamiltonian and super momentum constraints become
$$\eqalign{H~~ =~~ -~  &\left[{{F^{-1} M' R' + F \bP_R (\bP_M + \tau')}
\over {L}}\right]\cr &+~~  (\bP_\tau - M') \sqrt{1 + {\tau'}^2
/L^2}~~ =~~ 0\cr}\eqno(2.8)$$
and
$$H_r~~ =~~ M'\bP_M~  +~  R'\bP_R~  +~  \tau'\bP_\tau~~ =~~ 0, \eqno(2.9)$$
where we have used
$$L^2~~ =~~ F^{-1} R'^2~ -~ F(\bP_M + \tau')^2 \eqno(2.10)$$
and $F~~ =~~ 1-2M/R$. $L^2$, being the component $g_{rr}$ of the spherically 
symmetric metric in (2.2), must be positive definite everywhere. $F$ is positive 
in the exterior (Schwarzschild) region and negative in the interior and this will 
play an important role in the consistency conditions that follow.  By direct 
computation of Poisson brackets, it is easy to determine the ``velocities''  in 
terms of the conjugate momenta from the above expressions and they are 
$$\eqalign{\dtau~~ &=~~ N\sqrt{1+{\tau'}^2/L^2}~  +~  N^r \tau',\cr \dR~~ &=~~
- {{NF(\bP_M+\tau')}\over L}~  +~  N^r R',\cr \dM~~ &=~~ {{NR'\tau' (\bP_\tau-M')}
\over {L^3}}~  +~  N^r M'.\cr}\eqno(2.11)$$
The constraints in (2.8) and (2.9) generate the local symmetries of the theory. In the 
Schroedinger representation, they correspond to functional differential 
equations that the state functional must satisfy, i.e., the canonical variables must be 
raised to operator status and the constraints considered as operator constraints 
on the (Wheeler-DeWitt) wave functional $\psi_{WD}[\tau,R,M]$, 
$${\hat H} \psi_{WD}[\tau,R,M]~~ =~~ 0~~ =~~ {\hat H_r} \psi_{WD}[\tau,R,M].
\eqno(2.12)$$
These equations are, in fact, an infinite set of equations, one for each spatial point on
the spatial hypersurface.  

If the super momentum constraint in (2.9) is used to eliminate $\bP_M$ in the 
expression for the super Hamiltonian in (2.8), the latter  constraint turns 
into${}^{18}$
$$(\bP_\tau-M')^2~  +~  F \bP_R^2~  -~ {{M'}^2 \over F}~~ =~~ 0.\eqno(2.13)$$
We will now specialize to the black hole by requiring $M' = 0$ so that only the 
homogeneous mode of $M(t,r)$ survives.  In this way the dust is made tenuous 
and the proper time variable must be thought of as the proper time of a {\it test}
particle (an ideal clock) in free fall. To understand its role in the canonical 
reduction, recall that the Schwarzschild metric in comoving coordinates can be 
viewed as a special case of a marginally bound Tolman-Bondi metric for the 
collapse of inhomogeneous dust, which has the general form
$$ds^2~~ =~~ d\tau^2~ -~ {R'}^{2}(\tau,\rho)d\rho^2~ -~ R^2(\tau,\rho) 
d\Omega^2,\eqno(2.14)$$
where $\tau$ is the dust proper time and the prime denotes a derivative with 
respect to $\rho$. The curvature coordinate, $R(\tau,\rho)$, is obtained from 
Einstein's equations, after an appropriate scaling${}^{21}$, as
$$R^{3/2}~~ =~~ \rho^{3/2}~ -~ {3 \over 2} \sqrt{F(\rho)} \tau,
\eqno(2.15)$$ 
in terms of an arbitrary function (the ``mass'' function), $F(\rho)$. It represents 
the mass contained within a shell of radius $\rho$. The energy density of the 
dust is given by
$$\epsilon(\tau,\rho)~~ =~~ {{F'}\over {R^2 R'}}.\eqno(2.16)$$
In particular, if $F(\rho) = a^2$ (const.), the energy density vanishes everywhere 
except at the singularity, $R=0$, and the metric in (2.14) describes a (Schwarzschild) 
black hole of mass $a^2/2$. This follows from the coordinate transformation${}^{22}$
$$\eqalign{\tau~~ &=~~ T~ +~ a \int {{\sqrt{R}dR} \over {R-a^2}}\cr 
\rho^{3/2}~~ &=~~ R^{3/2}~ +~ {3 \over 2} a \left(T~ +~ a \int {{
\sqrt{R}dR} \over {R-a^2}}\right),\cr}\eqno(2.17)$$
which takes (2.14) into the standard Schwarzschild form, with $2M = a^2$ and where 
$T$ is the Killing time. This spacetime transformation may be re-expressed as a point 
transformation on the phase space.

The resulting Wheeler-DeWitt  equation (with $M'=0$) is decoupled, and  
the quantum state may be expressed formally as a direct product of states defined at 
each spatial point,
$$|\psi_{WD}\rangle~~ =~~ \prod_r |\Psi_r\rangle,\eqno(2.18)$$ 
where we have used $r$ as a label. Each of the $|\Psi_r\rangle$ is normalized 
 w.r.t. a suitable measure in the space of functions $\tau(r), R(r), M$  at fixed $r$
and the system reduces to a set of {\it independent}  Schroedinger equations, one 
for each spatial point. The wave functional is given by
$$\eqalign{\psi_{WD}[\tau,R,M]~~ &=~~  \prod_r \Psi_r[\tau(r),R(r),M]\cr
&=~~ \Psi_1 \otimes \Psi_2 \otimes ... \otimes ... \otimes \Psi_N.\cr}\eqno(2.19)$$
The second equation is written by imagining that a lattice is placed on each spatial 
hypersurface so that the classically continuous label coordinate, $r$, is discretized. 
In this form, the Wheeler-DeWitt wave functional represents a collection of, say, $N$ 
decoupled systems, each determined by the same Schroedinger equation and obeying the 
same boundary conditions. The precise value of $N$ cannot be determined at this 
level. We will ascertain its value by requiring it to maximize the 
density of states.

The second constraint enforces spatial diffeomorphism  invariance of  the wave 
functional on hypersurfaces orthogonal to the dust proper time. By taking functional 
derivatives of $\psi_{WD}$ in (2.19), it is easy to see that this constraint enforces 
$\Psi_r'[\tau(r),R(r),M]= 0$, where the prime denotes a derivative with respect to 
the label coordinate $r$. 

We also see, from eq. (2.13), that, for every label $r$, the (Schroedinger) equation 
reads
$$\nabla^2 \Psi~~ =~~ \gamma^{ab} \nabla_a \nabla_b \Psi~~ =~~ 
{\tilde H} \Psi~~ =~~ 0, \eqno(2.20)$$
where $\gamma_{ab}$ is the field space metric, $\gamma_{ab} = {\rm diag}(1,1/F)$, and 
$\nabla_a$ is the covariant derivative with respect to this metric.  The operator 
$\nabla^2$ is the Laplace-Beltrami operator having $\gamma_{ab}$ as its covariant metric, 
and eq. (2.20) is a massless ``Klein-Gordon'' equation.  It is
hyperbolic in the region $R < 2M$ (the interior of the Kruskal manifold) but 
elliptic in the region $R>2M$ (the exterior). This is because the quantity 
$F$ is negative in the interior, but positive in the exterior.  As shown in ref. [18], 
this means that the unique positive energy solution of the Wheeler-DeWitt equation
 in the exterior,  that is compatible with spatial diffeomorphism invariance, is 
identically zero. The dynamics is therefore confined to the {\it interior} of the 
hole.  This is consistent with the assumed geometry, as the asymptotic observer 
sees the exterior region of the spacetime as static.

In the interior, the Schroedinger equation is hyperbolic and it is convenient to 
transform to the coordinate $\bR_*$ defined by
$$\bR_*~~ =~~ -~ \sqrt{R(2M-R)}~ +~ M \tan^{-1}\left[{{R-M} \over {\sqrt{R(2M-R)}}}
\right]~. \eqno(2.21)$$
The new coordinate lies in the range $(-{{\pi M} \over 2},+{{\pi M} \over 2})$ 
and the wave equation, 
$$\partial_\tau^2 \Psi~ -~ \partial_*^2 \Psi~~ =~~ 0, \eqno(2.22)$$
now defines the quantum theory whose Hilbert space is ${\cal H} := {\cal L}^2
({\bf R}, d\bR_*)$ with inner product 
$$\langle \Psi_1,\Psi_2\rangle~~ =~~ \int_{-{{\pi M}\over 2}}^{+{{\pi M}\over 2}} 
d\bR_* \Psi_1^\dagger\Psi_2~ . \eqno(2.23)$$
The general (positive energy) solution is
$$\Psi_{in}~~ =~~ c_+(M) e^{-iE(\tau + \bR_*)}~~ +~~  c_-(M) e^{-iE(\tau - \bR_*)} 
\eqno(2.24)$$
where $c_\pm$ are functions only of $M$. We must impose the super momentum 
constraint, which reads
$$(\tau'+R') c_+(M) e^{-iE(\tau+\bR_*)}~~ +~~ (\tau'-R') c_-(M) e^{-iE(\tau-
\bR_*)}~~ =~~ 0, \eqno(2.25)$$
assuming $E>0$. A consistent and physically meaningful solution to this equation 
is $\tau' = \bR'_* = 0$. Returning to (2.11), we see that the choice implies that 
$\dtau = N$ and $\dM=0$. Setting $N=1$, the dust proper time turns into the 
asymptotic Minkowski time and the energy, $E$, should be associated with the 
ADM mass of the black hole.

Imposing continuity across the horizon, this solution will match the solution 
in the exterior ($\Psi = 0$) at $R=2M$, if
$$c_-(M)~~ =~~ -~  c_+(M)e^{-iEM\pi},\eqno(2.26)$$
so that the solution in the interior is now of the form${}^{18}$
$$\Psi_{in}~~ =~~ c_+(M) \left[ e^{-iE(\tau+\bR_*)}~~  -~~ e^{-iEM\pi} 
e^{-iE(\tau -\bR_*)}\right]~ .\eqno(2.27)$$ 
There does not seem to be a natural way to impose further boundary conditions. 
Certainly, boundary conditions cannot be imposed at the classical singularity 
where the canonical reduction will break down anyway. Nevertheless, one notes 
that the parity operator $\bR_* \rightarrow -\bR_*$ commutes with the 
Hamiltonian. States of definite parity will vanish at the classical singularity 
and exhibit a discrete spectrum given by
$$\eqalign{\Psi_{in}^{(+)}~~ &=~~ {1 \over {\sqrt{\pi M}}} e^{-iE\tau} 
\cos E\bR_*~~~~~~~~~~ EM~ =~ (2n+1)~ ,\cr \Psi_{in}^{(-)}~~ &=~~ {1 \over 
{\sqrt{\pi M}}} e^{-iE\tau} \sin E\bR_*~~~~~~~~~~ EM~ =~ 2n .\cr}
\eqno(2.29)$$
Furthermore, as the dust  proper time is identified with the asymptotic Minkowski 
time and the total energy  with the ADM mass of the black hole, we are led to the 
Bekenstein mass quantization rule
$$M_n~~ =~~ \sqrt{n} M_p\eqno(2.30)$$
for the definite parity states. 

States of indefinite parity do not admit a quantized mass spectrum, but there are 
three good reasons to confine attention to states of definite parity. Firstly, 
because the parity operator commutes with the hamiltonian, states of definite parity  
are guaranteed to remain so for all time, which is in harmony with our intuitive 
notion of an  ``eternal'' black hole.  Secondly, definite parity eigenstates  do 
not support the singularity at the origin, which, given that the entire canonical 
quantization program breaks down there, is an attractive feature. Finally, and 
perhaps most importantly, we shall count only the definite parity states in what 
follows and show that they fully account for the entropy of the black hole.
\vskip 0.25in

\noindent{\bf 3. The Entropy}

To compute the entropy, we must enumerate the states of the system. To this effect,
it is convenient to reformulate the problem by recognizing that the  wave equation 
at each label, $r$, in (2.20) is derivable from the action,
$$S~~ =~~ -{1\over 2} \sum_{r=1}^N \int_{\cal R} d^2 X\sqrt{|\gamma|} \gamma^{ab} 
\partial_a \Psi_r^\dagger \partial _b \Psi_r, \eqno(3.1)$$
where $X \in (\tau, \bR_*)$,  the integral is over the interior of the Kruskal 
manifold  and such that  $\Psi_r(\tau,\bR_*) = 0$ at $\bR_* = -\pi G M/ 2, +
\pi G M/ 2,~~ \forall~~ r$.  Imposition of these boundary conditions automatically confines  
attention to states of definite parity.  Recall that the Wheeler-DeWitt wave functional 
is a direct product state, as given in (2.19). As we will see, each component of the 
direct product can  be thought of as describing a tower of oscillators living in the 
internal space parametrized by the phase space coordinates $(\tau,\bR_*)$. 

Using (3.1), performing a mode expansion of $\Psi_r$ and combining both even and 
odd parities, we can express the contribution  of any one of the lattice sites to the total 
energy of the system in terms of pairs $\alpha_n,\beta_n$ of creation and annihilation 
operators as follows
$${\hat{\cal H}}_r~~ =~~ {{M_p^2} \over M} \sum_{n_r} (\alpha^\dagger_{n_r} \alpha_{n_r}~ +~ 
\beta^\dagger_{n_r} \beta_{n_r}),\eqno(3.2)$$
where
$$\eqalign{[\alpha_{n_r},\alpha^\dagger_{n_r}]~~ &=~~ n_r \cr [\beta_{n_r}, 
\beta^\dagger_{n_r}]~~  &=~~ n_r\cr}\eqno(3.3)$$
are the only non-vanishing commutators. At each site, $r$, one therefore has a hierarchy 
of two dimensional oscillators. The total  energy is the sum over contributions from each 
of the $N$ sites, i.e.,
$${\hat H}_{tot}~~ =~~ {{M_p^2} \over M} \sum_{r=1}^N \sum_{n_r} (\alpha^\dagger_{n_r} 
\alpha_{n_r}~+~ \beta^\dagger_{n_r} \beta_{n_r}),\eqno(3.4)$$
which gives (the total energy is the mass of the black hole)
$$\eqalign{M~~ &=~~ {1\over M} \sum_r m_r^2\cr &=~~ {{M_p^2} \over {M}}  
\sum_{r=1}^N \sum_{n_r,l_r} (n_r N_{n_r}~  +~ l_r K_{l_r})\cr \rightarrow~~ M~~ 
&=~~ {\sqrt{\cal N}} M_p,~~~~ {\cal N}~ \in~ {\bf  N} \cup \{0\}, \cr} \eqno(3.5) $$
where $n_r,l_r$ and $N_{n_r}, K_{l_r}$ are respectively the level number and the 
occupation number at level number $n_r(l_r)$, corresponding to the oscillators at site
$r$. 

Let $\rho_D(m_r)$ be the density of levels describing each site, $r$. This is just
the number of states with mass given by $m_r^2 = \nu_r M_p^2$ ($\nu_r \in {\bf N}$), 
and is known to have the asymptotic ($m_r >> M_p$) form${}^{23}$
$$\rho_D(m_r)~~ =~~ c \times m_r^{-(D+1)/2} \times \exp\left[2\pi\sqrt{D \over 6} {{m_r}
\over {M_p}} \right], \eqno(3.6) $$
where $c$ is a constant and $D$ is the dimension of the oscillator.  For a generic 
level density, $\rho_D(m_r)$, the density of states may be written as
$$\Omega(N,M)~~ =~~ \prod_{r=1}^N \int_{M_0}^\infty dm_r \rho_D(m_r) \delta
\left({1\over M} \sum_{s=1}^N m_r^2 - M\right),\eqno(3.7)$$
where $M_0$ is the lowest value of $m_r$ for which the density of levels is valid. As 
each site is fixed, there are no further phase space integrals. The delta function 
in (3.7) imposes energy conservation as required by (3.5).  Let us assume that 
$\rho_D(m_r)$ is such that the dominant contribution to the mass integrals comes 
from states with large mass (unless these states are forbidden by energy conservation). 
This is certainly true for the level density in (3.6). Then define the quantity
$$\sigma_N(M)~~ =~~ \int_{NM_0}^M dx \prod_{r=1}^N \int_{M_0}^\infty dm_r 
\rho_D(m_r)\delta({1\over M} \sum_r m_r^2 -x),\eqno(3.8)$$
in terms of which we may write 
$$\Omega(N,M)~~ =~~ {d \over {dM}} \sigma_N(M).\eqno(3.9)$$
The $\delta-$ function restricts the limits of the mass integrals in the 
definition of $\sigma_N(M)$, they no longer run to infinity. Following Carlitz${}^{17}$, 
we estimate the $r^{\rm th}$ integral by
$$\int_{M_0}^{\sqrt{M\Lambda_r(x)}} dm_r \rho_D(m_r),\eqno(3.10)$$
provided that the $\Lambda_r(x)$ are subject to the constraint
$$\sum_{r=1}^N \Lambda_r(x)~~ =~~ x~ . \eqno(3.11)$$
The maximum contribution to $\sigma_N(M)$ is obtained when the $\Lambda_r(x)$ are 
all of the order of $x/N$. This provides an estimate for the integrals in (3.8), and 
one finds, quite generally, 
$$\Omega(N,M)~~ =~~ a^N f(\xi)^N,\eqno(3.12)$$
where $\xi = b M^2/N M_p^2$, and $a$ and $b$ are dimensionless constants . The 
most probable number of sites is obtained by maximizing $\Omega(N,M)$ with 
respect to $N$. Therefore, consider
$${\partial \over {\partial N}} \ln \Omega(N,M)~~ =~~ \ln a~  +~ \ln f(\xi)~ -~ 
\xi {\partial\over {\partial \xi}} \ln f(\xi)~~ =~~ 0,\eqno(3.13)$$
which depends only on $\xi$. Assume that $\Omega$ is maximized for some 
value, say $\alpha^{-1}$, of $\xi$. Then this value gives the corresponding number 
of sites as
$$N_{max}~~ =~~ \alpha b {{M^2} \over {M_p^2}}. \eqno(3.14)$$
$N_{max}$ is therefore proportional to the area of the horizon. A remarkable 
consequence is that all the degrees of freedom can be {\it treated as though} they 
resided just there, that is on the horizon itself.

Equation (3.13) can be integrated and the solution written in terms of a single
dimensionless constant, $\gamma$. One finds that
$$e^{-\gamma\xi} f(\xi)~~ =~~ {1 \over a}.\eqno(3.15)$$
This constant, $\gamma$ then determines the maximum number of states according to 
$$\eqalign{\ln\Omega_{max} (M)~~ &=~~ N \ln (a)~ +~ N\ln f(\xi)|_{N_{\max}}\cr &=~~ 
N\ln e^{-\gamma \xi} f(\xi)~ +~ N\ln (a)~ +~ N\gamma\xi|_{N_{max}}\cr &=~~ 
N\gamma\xi|_{N_{max}}~~ =~~ b \gamma {{M^2} \over {M_p^2}}\cr}\eqno(3.16)$$
and is, itself, to be determined from (3.15) at $\xi = \alpha^{-1}$, {\it i.e.,}
$$\gamma~~ =~~ \alpha \ln [a f(\alpha^{-1})].\eqno(3.17)$$
One thus recovers the entropy
$$S~~ =~~ \ln\Omega_{max}(M)~~ =~~ {{b\gamma}\over {4\pi l_p^2}} \left({{\cal A} \over 
4}\right),\eqno(3.18)$$
where $\cal A$ is the horizon area. This is the area law, provided that $\gamma > 0$,
and it is independent of the precise form of the density of levels (except for 
proportionality constants), only requiring that the latter is such that the dominant 
contributions come from the most massive states permitted by energy conservation. $S$ 
inherits its dependence on $M$ only from the dispersion relation in (3.5).

As an example, let us use the density of levels as given in (3.6) although we will 
see that it is strictly not correct to do so. In the present context $D=2$, giving
$$\eqalign{\rho_2(m_r)~~ &=~~ c \times m_r^{-3/2} \times \exp\left[{2\pi\over{\sqrt 3}} 
{{m_r}\over {M_p}} \right]\cr \Omega(N,M)~~ &=~~ c^N \prod_{r=1}^N \int_{M_0}^\infty 
dm_r m_r^{-3/2} e^{2\pi m_r/{\sqrt{3}}M_p} \delta({1\over M} \sum_r m_r^2 -M).\cr} 
\eqno(3.19)$$
Performing all the necessary steps, one recovers (3.12) with
$$f(\xi)~~ =~~ \left[{\sqrt{\pi}} {\rm Erfi}(\xi^{1/4}) - {\sqrt{\pi}} {\rm Erfi}
(\xi_0^{1/4}) - {{e^{\xi^{1/2}}} \over {\xi^{1/4}}} +  {{e^{\xi_0^{1/2}}} \over 
{\xi_0^{1/4}}}\right], \eqno(3.20)$$
where ${\rm Erfi}(z) = {\rm Erf}(iz)/i$ is the imaginary error function,
$$\eqalign{\xi~~ &=~~ \left({{4\pi^2 M^2} \over {3NM_p^2}}\right)~~~~~ \rightarrow
~~ b~~ =~~ {{4\pi^2} \over 3}\cr \xi_0~~ &=~~ \left({{4\pi^2 M_0^2} 
\over {3M_p^2}}\right)\cr}\eqno(3.21)$$
and
$$a~~ =~~ \sqrt{{8\pi c^2} \over {\sqrt{3} M_p}}.\eqno(3.22)$$
Therefore one has,
$$S~~ =~~ {{\pi \gamma} \over {3 l_p^2}} \left({{\cal A} \over 4}\right).\eqno(3.23)$$
The constant $\gamma$ will depend on $M_0$, the lowest value of the mass for which 
the density of levels, $\rho_2(m_r)$, in (3.6) is valid, and on the constant, $c$. We have 
taken $c=\sqrt{M_p}$, $M_0 = M_p$ and found $\gamma \approx 0.089$.  $\gamma$ 
is found to decrease sharply with increasing $M_0/M_p$. This behavior contrasts 
with the expected value of $\pi\gamma/3 \approx 1$. The reason for this 
discrepancy may be traced to our use of the {\it asymptotic} level density. Recall 
that  $m_r^2  \leq M^2/N$ and we found, quite generally, that the density 
of states was maximized when $N \sim M^2/M_p^2$. This implies that $m_r \sim M_p$, 
which contradicts our use of (3.6). 

When only the lowest levels are occupied, we can approximate $\rho_D(m_r)$ by a constant, 
say $k/M_p$. This gives $f(\xi) = \xi^{1/2}$, $b = k^2$ and $a=1$, and inserting these 
values into (3.17) and (3.18) yields
$$S~~ =~~ {{k^2} \over {8\pi e l_p^2}} \left({{\cal A} \over 4}
\right). \eqno(3.24)$$ 
To recover Hawking's temperature we must take $k \sim \sqrt{8\pi e}$.
\vskip 0.25in

\noindent{\bf 4. Discussion}

We have shown that the canonical reduction of spherical geometries due to 
Kucha\v r${}^{19}$, combined with the introduction of an extrinsic time variable via 
the coupling to non-rotating dust as proposed by Kucha\v r and Brown${}^{20}$, leads 
to a remarkably simple description of the eternal Schwarzschild black hole. The 
Wheeler-DeWitt wave functional was seen to be expressible as a direct product state, 
which was interpreted as a collection of $N$ oscillator hierarchies, each of 
which could be described by a free, massless, complex scalar propagating in a ``flat'' two 
dimensional background (the internal, ``metric'' space) and confined to the interior 
of the Kruskal manifold. The canonical quantization program does not allow for an estimate 
of $N$, but, using techniques from  the study of the statistical behavior of  dual models, 
the Veneziano model and the statistical bootstrap model by Frautschi${}^{16}$ 
and Carlitz${}^{17}$, we discovered that the value of $N$ that maximizes 
the density of states is given by $N \sim M^2/M_p^2$, i.e., the number of (lattice) 
sites is proportional to the area of the horizon, implying that all the degrees of 
freedom may be thought of as residing on the horizon itself. It is a remarkable result 
that probably has a deeper meaning. One is tempted to speculate, for example, that this,
or some similar mechanism, may be a consequence of, or indeed a justification for, some 
form of a ``holographic'' principle. Moreover, because the total energy, $M$, of the 
black hole, is divided between these $N$ lattice sites according to the dispersion 
relation $\sum_r m_r^2 = M^2$, each oscillator is virtually in its ground state, with a 
small associated degeneracy. The picture that emerges thus coincides also with 
Bekenstein's original way${}^{3}$ of estimating the entropy by dividing the horizon 
into cells of Planck area, each of which has a small number of associated states.

We were able to calculate the statistical entropy of  the black hole by evaluating 
the microcanonical density of states of the system. Thus we recovered  the area law 
of black hole thermodynamics, which is seen to be the consequence of the dispersion
relation in (3.5) and the area quantization rule.  It is noteworthy that all 
the states are explicitly displayed, both the mass quantization rules and the statistical 
entropy are recovered, and no explicit appeal to boundary states has had to be made in 
this approach. 
\vskip 0.25in

\noindent{\bf Acknowledgements:} 

\noindent The author acknowledges the partial support of FCT, Portugal, under 
contracts number FMRH/BSAB/54/98 and number CERN/P/FAE/1172/97.
\vfill\eject
\noindent{\bf References:}

\item{1.}S. W. Hawking, Phys. Rev. Letts. {\bf 26} (1971) 1344.

\item{2.}S. W. Hawking, Comm. Math. Phys. {\bf 43} (1975) 199.

\item{3.}J. D. Bekenstein, Lett. Nuov. Cim. {\bf 11} (1974) 467; ``Black Holes: 
Classical Properties, Thermodynamics and Heuristic Quantization'', in the IX 
Brazilian School on Cosmology and Gravitation, Rio de Janeiro 7-8/98, 
gr-qc/9808028; ``Quantum Black Holes as Atoms'', in the VIII Marcel Grossmann 
Meeting on General Relativity, Jerusalem, June 1997, gr-qc/9710076; Phys. 
Lett. {\bf B360} (1995) 7; Phys. Rev. Lett. {\bf 70} (1993) 3680.

\item{4.} D. Christodoulou and R. Ruffini, Phys. Rev. {\bf D4} (1971) 3552.

\item{5.} M. Bojowald, H.A. Kastrup, {\it  The Area Operator in the Spherically 
Symmetric Sector of Loop Quantum Gravity}, hep-th/9907043; K. V. Krasnov, Class. 
Quant.Grav. 16 (1999) L15; Gen. Rel. Grav. 30 (1998) 53; A. Ashtekar and K. V. 
Krasnov, {\it Quantum Geometry of Black Holes}, gr-qc/9804039; A. Ashtekar, J. Baez, 
A. Corichi, K Krasnov, Phys. Rev. Letts. {\bf 80} (1998) 904; A. Ashtekar, J. Lewandowski, 
Class. Quant. Grav. {\bf 14} (1997) A55; J. Louko and J. M\"akel\"a, Phys. Rev. {\bf 
D54} (1996) 4982; J. Louko and S. Winters-Hilt, Phys. Rev. {\bf D54} (1996) 2647; J. 
M\"akel\"a, Phys. Letts. {\bf B390} (1997) 115; H. A. Kastrup, Phys. Lett. {\bf B385} 
(1996) 75; Phys. Letts. {\bf B413} (1997) 267; Phys. Letts. {\bf B419} (1998) 40; Y. 
Peleg, Phys. Lett. {\bf B356} (1995) 462.

\item{6.}A. W. Peet, Class. Quant. Grav. {\bf 15} (1998) 3291; K. Sfetsos, 
K. Skenderis, Nucl. Phys. {\bf B517} (1998) 179; A. Strominger and C. Vafa, 
Phys. Lett. {\bf B379} (1996) 99;  C. O. Lousto, Phys. Rev. {\bf D51} (1995) 
1733; M. Maggiore, Nucl. Phys. {\bf B429}  (1994) 205;  Ya. I. Kogan, JETP
Lett. {\bf 44} (1986) 267.

\item{7.}For a review of these and other approaches see, for example, S. Mukohyama, 
``The Origin of Black Hole Entropy'', gr-qc/9812079.

\item{8.}A. Strominger, JHEP 9802 (1998) 009; J. M. Maldacena and A. Strominger,
JHEP 9802 (1998) 014.

\item{9.}M. Ba\~nados, C. Teitelboim, J. Zanelli, Phys. Rev. Letts. {\bf 69}  (1992)
1849. 

\item{10.}J. D. Brown, and M. Henneaux, Comm. Math. Phys. {\bf 104} (1986) 207.

\item{11.}J. A. Cardy, Nucl. Phys. {\bf B270} (1986) 186.

\item{12.}S. Carlip, Phys. Rev. Letts, {\bf 82} (1999) 2828; Phys. Rev. {\bf D51}  
(1995) 632.

\item{13.}K. V. Krasnov, Gen. Rel. Grav. 30 (1998) 53; Phys. Rev. {\bf D55} (1997) 
3505.

\item{14.}C. Rovelli, Phys. Rev. Letts. {\bf 14} (1996) 3288; Helv. Phys. Acta {\bf 69} 
(1996) 582.

\item{15.}A. Ashtekar, J. Baez, A. Corichi, K. Krasnov, Phys. Rev. Letts. {\bf 80} 
(1998) 904; A. Mamen, Phys. Letts. {\bf B394} (1997) 269; A.P. Balachandran, L. Chandar 
and A. Mamen, Int. J. Mod. Phys. {\bf A12} (1997) 625; Nucl. Phys. {\bf B461} (1996) 
581.

\item{16.}S. Frautschi, Phys. Rev. {\bf D3}, (1971) 2821.

\item{17.}R. D. Carlitz, Phys. Rev. {\bf D5} (1972) 3231.

\item{18.}Cenalo Vaz and Louis Witten, Phys. Rev. {\bf D60} (1999) 024009.

\item{19.}K. V. Kucha\v r, Phys. Rev. {\bf D50} (1994) 3961.

\item{20.}J. D. Brown and K. V. Kucha\v r , Phys. Rev. {\bf D51} (1995) 5600.

\item{21.}see, for instance, P.S. Joshi and T.P. Singh, Phys. Rev. {\bf D51}
(1995) 6778 and refs. therein.

\item{22.}L.D. Landau and E.M. Lifshitz, {\it The Classical Theory of 
Fields}, Butterworth-Heinemann (1997).

\item{23.}K. Huang and S. Weinberg, Phys. Rev. Letts. {\bf 25}  (1970) 895, generalized 
the original work of Hardy and Ramanujan, Proc. Lond. Math. Soc. {\bf 17} (1918) 75. 
Applications to superstring theory can be found in M.B. Green, J.H. Schwartz and 
E. Witten, {\it Superstring Theory}, Vols. I and II, Cambridge University Press,  New 
York  (1987).
\bye